\documentstyle[12pt]{article}
\begin{document}
\def\thebibliography#1{\section*{REFERENCES\markboth
 {REFERENCES}{REFERENCES}}\list
 {[\arabic{enumi}]}{\settowidth\labelwidth{[#1]}\leftmargin\labelwidth
 \advance\leftmargin\labelsep
 \usecounter{enumi}}
 \def\newblock{\hskip .11em plus .33em minus -.07em}
 \sloppy
 \sfcode`\.=1000\relax}
\let\endthebibliography=\endlist

\hoffset = -1truecm
\voffset = -2truecm

\title{\large\bf
A Dynamical Principle For 3D-4D Interlinkage In Salpeter-like Equations  
}
\author{
{\normalsize \bf
A.N.Mitra \thanks{e.mail: ganmitra@nde.vsnl.net.in} \quad and B.M.Sodermark*
}\\
\normalsize  244 Tagore Park, Delhi-110009, India \\
\normalsize *High Energy Lab, Dept of Phys, U of Delhi, Delhi-7 India
} 
\date{28 February 2001}
\maketitle


\begin{abstract}

The half-century old Markov-Yukawa Transversality Principle 
($MYTP$) which provides a theoretical rationale for the covariant
instantaneous approximation ($CIA$) that underlies all Salpeter-like 
equations, is generalized to a covariant null-plane ansatz 
($CNPA$). A common characteristic of both formulations is an exact 
3D-4D interlinkage of BS amplitudes which facilitates a two-tier
description: the 3D form for spectroscopy, and the 4D form for
transition amplitudes as 4D loop integrals. Some basic applications
of $MYTP$ on the covariant null plane (quark mass function, vacuum
condensates, and decay constants) are given on the lines of earlier
applications to these processes under $CIA$.   
   


\end{abstract}

\section*{1  Introduction}
   
For a {\it relativistic} 2-body problem, the historical issue of 3D 
reduction from a 4D BSE has been in the forefront of its physics from the 
outset: Instantaneous approximation [1]; Quasi-potential approach [2]; 
variants of on-shellness of propagators [3]. This is a sort of recognition
of the intractability of the strong interaction problem which had led
Bethe as early as in the Fifties to invoke his famous Second Principle
Theory, signifying the postulation of an effective $N-N$ interaction for 
a microscopic understanding of the physics of nuclei. The Bethe-Salpeter 
Equation (BSE) is a relativistic version of this Principle, initially
at the nucleon-nucleon level, later adapted to the quark level. Now
one might ask: why 3D reduction at all ? One possible answer is the 
need to preserve the $probability$ interpretation which is unavailing 
in its 4D form since the BSE is only an $approximate$  description 
(in the `ladder approximation') which stems from an effective 4-fermion 
Lagrangian mediated  by, say, a gluonic propagator which serves as the 
$kernel$ of the BSE in the lowest order [4]. This may be contrasted with 
the Schwinger-Dyson Equations (SDE) which are an infinite chain  of  
equations connecting successively higher order vertex functions [5] that
stem from an $exact$ Lagrangian which characterizes QED or QCD.  
\par
	The  usual 3D reduction methods [1-3] have one feature in common: 
The starting BSE is 4D in all details, {\it {including its kernel}}, but 
the associated propagators are manipulated in various ways to reduce the 4D
BSE to a 3D form as a fresh starting point, giving up its original 4D form.
An alternative approach which was pioneered by Weinberg [6], is 
intrinsically 3D in character (analogous to the Tamm-Dancoff method [7]).
It was refined, among others, by Kadychevsky [8a] and Karmanov in a 
covariant light-front style [8b], and reviewed by the Grenoble group [9]. 
\par
	An alternative approach to 3D reduction of more recent origin 
[10,11] is based on the Markov-Yukawa Transversality Principle ($MYTP$) 
[12], with  a Lorentz-covariant 3D support postulated at the outset for 
the pairwise BSE kernel $K$ by demanding that it be a function of only 
${\hat q}_\mu = q_\mu-q.PP_\mu/P^2$, so that ${\hat q}.P \equiv 0$; but 
the propagators are left untouched in their original 4D forms.  Now
unlike the traditional methods [1-3] which give only a one-way connection,
($4D\rightarrow 3D$), $MYTP$ [12] allows a two-way $interconnection$ [11] 
between the 4D and 3D BSE forms, the latter rooted in spectroscopy [13]. 

\subsection*{1.1  $MYTP$ and 3D-4D Interlinkage of BS Amplitudes}

Our main concern in this paper is with this `alternative approach'
epitomised by the $MYTP$ [12] which provides a rationale for the 
Instantaneous Approximation insofar as the latter also amounts to a 3D
support to the BS kernel [10]. By the same logic, the original Salpeter 
Equation [1] which stems from the adiabatic approximation to the BSE [14], 
is also amenable to $MYTP$ [12,15], except for the apparent loss of 
covariance which is a mere technicality [10]. We shall find 
it convenient in this paper to speak of all such BSE's with 3D support 
to their respective kernels as `Salpeter-like' equations, whether at the 
atomic level [16] or at the quark-hadron level [17]. 
\par
	 Since $MYTP$ [12] is characterized by the field dependence on both coordinate and momentum, it violates local micro-causality (a basic requisite for the 
theory of elementary particles). QCD [18] changed this perspective by pushing the status of hadrons from the elementary to a $composite$ level, characterized by
bilocal fields [19]. Within such a bilocal scenario, the total 4-momentum $P_\mu$ of the composite hadron provides a naturally preferred direction which forms the 
basis for a covariant 3D support to the interaction kernel [10-11]. For a 
bilocal field ${\cal M}(z,X)$ [19], the  Transversality condition on the 
BS kernel was shown [20] to be equivalent to a `gauge principle'  which 
expresses the redundance of the $longitudinal$ component of the  relative 
momentum for the physical interaction between the two constituents. This in turn  
suffices [10] to show that the 3D Salpeter equation [1] is an exact  consequence of 
the covariant 3D support to the Bethe-Salpeter kernel, with $P_\mu$ as the 
preferred direction, thereby giving a formal basis, not only  to the 3D Salpeter equation, but also to its reconstructed 4D form [11]. The same logic of 
course goes through for the spinor case too [15].     
\par
	Now the reconstructability of 4D BS amplitudes in terms of 3D
ingredients had been noticed empirically [21] within the 
instantaneous approximation  to a QCD motivated BSE framework [22], and
applied to spectroscopy and processes [23]. Subsequently the Bonn Group
[24], studying Salpeter-like equations with t'Hooft instantons [25],
also noticed the same property. The more important issue, apparently 
not addressed in these approaches [21,24], was one of support from a 
deeper underlying theory. This is now provided by $MYTP$ [12], with a 
gauge covariant meaning [20] to the Instantaneous Approximation ($CIA$) 
that characterizes all Salpeter-like equations. The 3D-4D interlinkage of Bethe-Salpeter amplitudes  was recently generalized to the $qqq$ problem [26].

\subsubsection*{Physical Ingredients for $MYTP$ Based BSE}
 
The $MYTP$-governed BSE  of course needs supplementing by physical 
ingredients to define a BSE kernel, much as 
a Hamiltonian needs a properly defined `potential'. However its canvas
is broad enough to accommodate a wide variety of kernels which must in turn 
be governed by independent physical principles. In this respect, the orthodox 
view (which we adopt) is to keep close to the traditional 4D BSE-cum-SDE 
methods [27] which is a space-time extended version of NJL's [28] 
Dynamical Breaking of Chiral Symmetry ($DB{\chi}S$) for 4-quark interaction 
via vector exchange [27]. This generates a mass-function $m(p)$ via 
Schwinger-Dyson equation (SDE) [5], which accounts for the bulk of the 
constituent mass of $ud$ quarks via Politzer additivity [29]. Indeed, the 
BSE-SDE formalism [27] can be simply adapted [30] to the $MYTP$ form [4]
which gives 3D spectra of both hadron types [31] under a common 
parametrization for the gluon propagator, as well as a self-consistent SDE 
determination [30] of the constituent mass. 

\subsection*{1.2  $MYTP$ via Covariant Null-Plane Ansatz ($CNPA$)}

Despite these attractions, the $CIA$ formulation of $MYTP$ gives rise to ill-defined  
4D loop integrals due to a `Lorentz-mismatch' among the rest-frames of the participating hadronic composites, resulting in time-like momentum components 
in the (gaussian) factors associated with their vertex functions. This is 
especially so for triangle loops and above, such as the pion form factor and $\rho-\pi\pi$ coupling  where this disease shows up as unwarranted "complexities" 
[32] in the amplitudes, while one- and two-quark loops [33] just escape this pathology. A possible remedy, without giving up the obvious advantage of $MYTP$  
for a 3D-4D interconnection, is a {\it light-front}/ ({\it null-plane}) 
formulation a la Dirac [34] by virtue of its bigger $(7)$ stability group  
compared with $6$ for the {\it {instant form}}($CIA$) theory. The Dirac-Weinberg theory developed into a covariant LF dynamics [35,9, 36] from a non-covariant
formulation [37,38] in null-plane variables. To make this language accessible
to $MYTP$ it is only necessary to generalize it from $CIA$  to a {\it {covariantly defined}} null-plane ansatz ($CNPA$) which has the potential to cure the `Lorentz mismatch' [32] disease. A recent calculation of the pion form  factor, using quark triangle loops [39] suggested that this is indeed possible, so that it makes sense 
to systematise the $CNPA$ formulation on closely parallel lines to $CIA$ 
[30, 33], through a prior calibration to some  standard physical processes.  
  
\subsection*{1.3   Objective and Scope of the Paper}

In this paper, we seek to  generalize $MYTP$ as an ansatz on the 
covariant null plane ($CNPA$), for a treatment of Salpeter-like equations. 
To bring out its close similarity with the (earlier) $CIA$ formulation, 
we provide a summary background in Appendix A, recalling the gauge 
basis [20] of $MYTP$ [12], and recapitulating the main steps under $CIA$ 
for an exact 3D-4D interconnection  between the  
corresponding BS amplitudes for spinless quarks. Appendix B gives a
corresponding derivation for the (fermionic) Salpeter equation [10,24,15].  
\par
	With this background, we formulate $MYTP$ [12] on a covariant null-plane 
in Section 2 by demanding the BSE kernel $K$ for pairwise interaction to be a function  of relative momentum ${\hat q}$ which is $transverse$ to the composite 4-momentum $P_\mu$, on the lines of $CIA$, but now the third component of 
${\hat q}$ must be suitably defined [39] on the Covariant null-plane as to be $independent$ of the time-like components of $q$ at $all$ the hadron-quark 
vertices of any loop. This in general makes ${\hat q}$  dependent 
on the orientation $n_\mu$ of the null-plane but this turns out 
to be merely a technical formality. The 3D-4D interconnection of
BS amplitudes follows in close analogy to the $CIA$ case [11].
\par
	Sect.3 describes the realistic case of fermion quarks, in which
the $CNPA$ formulation is made with a Gordon reduced modification of the 
BSE [22], again in close analogy to $CIA$.  The evaluation is 
greatly simplified by the observation that the reduced 3D BSE under 
$CNPA$  is algebraically equivalent to the corresponding $CIA$
form [30], so that the $CNPA$ extension works with the $same$
parametrization as $CIA$ [30] which is attuned to spectroscopy [31].    
However the reconstructed 4D vertex function in $CNPA$ is different from $CIA$. 
The techniques of $CNPA$ are illustrated with the typical example of
$\pi \rightarrow 2\gamma$, with normalizations sketched in Appendix C
for both types ($P,V$) of $q{\bar q}$ hadrons. Sections 4 and 5 give 
a $CNPA$ derivation for two key physical parameters i) quark mass function 
and ii) $q{\bar q}$ vacuum condensates, on the lines of $CIA$ [30],together 
with a brief comparison with perturbative QCD. Further, in view of the fundamental
 nature of the electroweak decay constants $f_P$ and $g_V$, Appendix D 
collects a quick $CNPA$ derivation of these quantities for completeness.
Sect.6 concludes with a resume and a critical comparison of Salpeter-like
equations with the more conventional (4D) BSE-SDE types.

\section*{2 \quad Salpeter-like Equations on Covariant Null Plane}

Since the central theme of the paper concerns a generalization 
of $MYTP$ on the Covariant null-plane (light-front), with a view to
expand its applicational base,we start by defining a 3D support to the 
BS kernel on the light front/null plane, on the lines of $CIA$. Now   
a covariant null-plane orientation may be represented by the 4-vector 
$n_\mu$, as well as its dual ${\tilde n}_\mu$, obeying the normalizations 
$n^2 = {\tilde n}^2 =0$ and $n.{\tilde n} = 1$. In the standard null-plane
(euclidean) notation, these quantities are $n=(001;-i)/\sqrt{2}$ and 
${\tilde n}=(001;i)/\sqrt{2}$, while the two transverse directions are 
denoted by the subscript $\perp$ on the concerned momenta. The $n$-dependence 
of various momenta ensures explicit covariance, whose notation is 
normalized to the standard null-plane notation $p_{\pm} = p_0 \pm p_3$,  
as  $p_+ = n.p \sqrt{2}$; $p_- = -{\tilde n}.p \sqrt{2}$, while the 
$perp$-components continue to be denoted by $p_{\perp}$ in both notations.

\setcounter{equation}{0}
\renewcommand{\theequation}{2.\arabic{equation}}
 
For the various quantities (masses, momenta, etc) we stick to the 
notation of [11], (see Appendix A), except when new features arise.  
For the relative momentum $q={\hat m}_2 p_1-{\hat m}_1 p_2$, where 
$P=p_1+p_2$ is the total 4-momentum of the hadron, the component playing 
the null-plane analogue of $P.qP/P^2$ in the instant form [11], now
needs to be more carefully defined so that the time-like component
does $not$ implicitly appear; for it is this `third component' that
causes the `Lorentz mismatch' disease by bringing in time-like
components via Lorentz transformations among different vertex functions.
With a little trial and error, the desired quantity turns out to be [39] 
$$ q_{3\mu} = x P_n n_\mu ; \quad P_n=P.{\tilde n}; \quad x=n.q/n.P$$,
giving ${\hat q}^2$ = $q_\perp^2 + z^2M^2$, as a check. We now collect 
the following definitions/results which will be freely used in this paper:
\begin{eqnarray}\label{2.1}
 q_\perp &=& q-q_n n; \quad {\hat q}=q_\perp+ x P_n n; \quad x=q.n/P.n;
 \quad P^2 = -M^2; \\ \nonumber
 q_n,P_n &=& {\tilde n}.(q,P);{\hat q}.n = q.n; \quad {\hat q}.{\tilde n} = 0; 
\quad P_\perp.q_\perp = 0;   \\ \nonumber 
     P.q &=& P_n q.n + P.n q_n; \quad P.{\hat q} = P_n q.n; \quad
{\hat q}^2 = q_\perp^2 + M^2 x^2
\end{eqnarray}        
\par
	To fix the ideas,we first consider the case of spinless quarks,to be 
followed by the more realistic case of fermion quarks in Section 3.

\subsection*{ 2.1 3D-4D BSE on Cov. Null Plane : Spinless Quarks}
 
Our first task is to derive the reduced 3D BSE (wave-function $\phi$) from 
the 4D BSE with spinless quarks (wave-function $\Phi$) when its kernel $K$ is 
{\it decreed} to be independent of the component $q_n$, i.e., 
$K=K({\hat q},{\hat q}')$, with ${\hat q}$ = $(q_\perp, x P_n n)$ [39], 
in accordance with the MYTP [12] condition imposed on the light front. 
The 4D BSE with such a kernel is, c.f., eq.(A.2):       
\begin{equation}\label{2.2}
i(2\pi)^4 \Phi(q) = {\Delta_1}^{-1} {\Delta_2}^{-1}  \int d^4q' 
K({\hat q},{\hat q'}) \Phi(q')
\end{equation}
where $m_i$ is the mass of quark $\#i$,  
$$\Delta_i = {p_i}^2 + {m_i}^2; \quad  P^2 = -M^2 $$
and $d^4 q$ = $d^2 q_\perp dq_3 dq_n$. Now define a 3D wave function

$$\phi({\hat q}) =  \int  d{q_n}\Phi(q)$$ 

and use this result on the RHS of (2.2) to give 
\begin{equation}\label{2.3}
i(2\pi)^4 \Phi(q) = {\Delta_1}^{-1} {\Delta_2}^{-1} 
\int d^3 q' K({\hat q},{\hat q}')\phi({\hat q}')  
\end{equation}
Now integrate both sides of eq.(2.3) w.r.t. $dq_n$ to give a 3D BSE 
in the variable ${\hat q}_\mu$:
\begin{equation}\label{2.4}  
(2\pi)^3 D_n({\hat q}) \phi({\hat q}) =  \int d^2{q_\perp}'d{q_3}'  
K({\hat q},{\hat q}') \phi({\hat q}')
\end{equation}
where the function $D_n(\hat q)$, is defined as in (A.7) for $CIA$ [11]: 
\begin{equation}\label{2.5}
\int d{q_n}{\Delta_1}^{-1} {\Delta_2}^{-1} = 2{\pi}i D_n^{-1}({\hat q})
\end{equation}
and may be obtained by standard null-plane techniques [37,23]  as follows. 
In the $q_n$ plane, the poles of $\Delta_{1,2}$ lie on opposite sides of 
the real axis, so that only $one$ pole will contribute at a time. 
Taking the $\Delta_2$-pole, which gives 
\begin{equation}\label{2.6}
2q_n = -{\sqrt 2} q_- = {[m_2^2 + (q_\perp-{\hat m}_2P)^2]} /
 {({\hat m}_2 P.n - q.n)}
\end{equation}
the residue of $\Delta_1$ works out from (2.1) as $2P.q = 2P.n q_n+2P_n q.n$, 
where a `collinearity frame' $P_\perp.q_\perp = 0$ [39] has been 
(temporarily) employed to simplify the calculations. And when the value 
(2.6) of $q_n$ is put in (2.5), one obtains (with $P_n P.n = -M^2/2$): 
\begin{equation}\label{2.7}
D_n({\hat q}) = 2P.n ({\hat q}^2 -\frac{\lambda(M^2, m_1^2, m_2^2)}{4M^2}); 
\quad  {\hat q}^2 = q_\perp^2 + M^2 x^2; \quad x = q.n/P.n      
\end{equation}
Now a comparison of (2.2) with (2.4) relates the 4D and 3D wave-fns: 
\begin{equation}\label{2.8}  
2{\pi}i \Phi(q,P) = D_n({\hat q}){\Delta_1}^{-1}{\Delta_2}^{-1} \phi({\hat q})
\end{equation}
which is valid near the bound state pole. The BS vertex function now becomes 
$\Gamma = D_n \times \phi/(2{\pi}i)$, just as in eq.(A.9) for $CIA$. This
result is formally {\it covariant}, albeit $n_\mu$-dependent, yet agrees
with the (apparently non-covariant) null-plane result [23] for $D_+$.    

\section*{3  Fermion Quarks: Full BSE Structure} 

\setcounter{equation}{0}
\renewcommand{\theequation}{3.\arabic{equation}}                  
             	
We now come to the more realistic case of fermion quarks within the 
SDE-BSE framework born out of an $effective$ gluon-exchange mediated 
4-fermion coupling at the input Lagrangian level with `current' (almost 
massless) quarks. The gluonic propagator encompasses both the 
perturbative and non-perturbative regimes, and automatically preserves 
the chiral character of the input Lagrangian [27, 30] which is broken a 
la $DB{\chi}S$ [28] in the solution of the corresponding SDE [27],
albeit with $MYTP$ constraints [4,30]. This step generates the 
dynamical mass function $m(p)$ [27,30] whose low momentum limit $m(0)$ 
gives the bulk contribution to the {\it constituent} mass $m_{cons}$, 
while the {\it current} mass $m_{curr}$ for $uds$ quarks (that enter 
the input Lagrangian) gives a small effect. This last is in keeping 
with Politzer's  Additivity principle [29], viz., $m_{cons}$ = 
$m_{curr}+m(0)$, which provides a rationale for the quark masses usually 
employed in potential models. The only extra ingredient to be incorporated 
in this formalism is the $MYTP$ constraint on the effective 4-fermion 
interaction to have a covariant 3D support [4,30]. The appropriate 
gluon propagator between two fermion quarks which meets this requirement,
must be taken in a covariant fashion [30]. 
\par
	 Next we write down the $MYTP$ governed BSE structure for fermion 
quarks with $CIA$-like support to its kernel [30], but now  under $CNPA$ :
\begin{equation}\label{3.1}  
i(2\pi)^4 \Psi(q,P) = S_{F1}(p_1) S_{F2}(p_2) \int d^4 q' 
K({\hat q}, {\hat q}') \Psi(q',P); \quad K= F_{12} i\gamma_\mu^{(1)}
i\gamma_\mu^{(2)} V({\hat q},{\hat q}')
\end{equation}
where $F_{12}$ is the color factor $\lambda_1.\lambda_2/4$ and the $V$- 
function expresses the scalar structure of the gluon propagator in the 
perturbative (o.g.e.) plus non-perturbative regimes. The hat notation ${\hat q}_3$
for $CNPA$ implies that the longitudinal component is now $q_{3\mu} = x P_n n_\mu$, with  $P_n$=$P.{\tilde n}$. (Note that in a $CIA$ formulation [11,30],the corresponding longitudinal component $cannot$ be defined so expicitly, as it gets `mixed up' with the `scalar' (time-like) component; this is the source of 
ill-defined gaussian integrals [32] under $CIA$). Now the full structure of $V$ under $MYTP$ has considerable flexibility, since the only constraint is one of
transversality of the relative momentum $q$ to the total momentum
$P$. Nevertheless it is instructive to list a concrete form [23,31]  as a 
prototype of the dynamics of inter-connection of the 4D amplitudes with the 3D
spectroscopy in actual practice. [The interested reader should 
have no difficulty in setting up his own version of $V$ as long it
conforms to the $MYTP$ constraints]. Using the simplified notations 
$k$ for $q-q'$, and $V({\hat k})$ for the $V$ function, we have: 
\begin{equation}\label{3.2}
V({\hat k})=4\pi\alpha_s/{\hat k}^2+{3 \over 4}\omega_{q{\bar q}}^2 
 \int d{\bf r} [r^2 {(1+4A_0 {\hat m}_1{\hat m}_2 {M_>}^2 r^2)}^{-1/2}
 -C_0/\omega_0^2] e^{i{\hat k}.{\bf r}};
\end{equation}
\begin{equation}\label{3.3}
\omega_{q{\bar q}}^2 = 4M_> {\hat m}_1 {\hat m}_2 \omega_0^2 \alpha_s({M_>}^2);
\quad \alpha_s(Q^2)={{6\pi} \over {33-2n_f}} {\ln (M_>/\Lambda)}^{-1};
\end{equation}
\begin{equation}\label{3.4}
{\hat m}_{1,2} = [1 \pm (m_1^2 - m_2^2)/M^2]/2; \quad 
M_> = Max (M, m_1+m_2); \quad C_0 = 0.27;\quad A_0=0.0283  
\end{equation}
And the values of the basic constants (all in $MeV$) are [23,31]
\begin{equation}\label{3.5}
\omega_0= 158 ;\quad m_{ud}=265 ;\quad m_s=415 ; \quad m_c=1530 ; 
\quad m_b = 4900.
\end{equation}     

\subsection*{3.1   ``Off-Shell'' Gordon Reduction}
 
The BSE form (3.1) is unfortunately not the most convenient  for wider
applications in practice, since the Dirac matrices entail several coupled 
integral equations. Indeed, it was noticed at an early stage of the BSE
programme [22] (independently of $MYTP$!) that a considerable simplification is 
effected by expressing them in `Gordon-reduced' form, (permissible on the mass 
shells, or better on the surface $P.q = 0$ [23]), a step which may be 
regarded as a sort of `analytic continuation' of the $\gamma$- matrices to 
`off-shell' regions (i.e., away from the surface $P.q =0$). Admittedly 
this constitutes a conscious departure from the original BSE structure 
(3.1), but such technical modifications are not unknown in the BSE 
literature [40] in the interest of greater manoeuvreability, without 
giving up the essentials. Such a step is not unreasonable, in view 
of the  "effective" nature of the BS kernel. Moreover, the effect of 
this step can be strictly monitored, since the neglected effects 
may still be kept track of by treating the $difference$ of the $exact$ 
and the `Gordon-reduced' kernels as a perturbation. On the other hand, 
the advantages of Gordon reduction are substantial, since it cures in a 
single stroke, a very troublesome disease which is known in the literature 
as a`continuum dissolution disease', first noted half a century ago [41], 
but revived in more recent times in the context of a `Volks Theorem' [42] 
concerning the mixing of positive an negative states that is inherent in 
a relativistic dynamcs, which tends to produce an unrenormalizable wave 
function in an $n-body$ system, where $n \geq 3$, while $n=2$ just 
escapes this pathology.              
\par
	These arguments form the basis of the suggestion [22,43] for a
`Gordon-reduced' form for the BSE (3.1), which stems in the first place
from an $effective$ 4-fermion interaction [4] at the Lagrangian level. 
To link up the Gordon-reduced fermion BSE structure with the `scalar' form
in Section 2, we first define an auxiliary function $\Phi(q,P)$ connected 
with  $\Psi(q,P)$ as follows [43]:
\begin{equation}\label{3.6} 
\Psi(q,P) = (m_1-i\gamma^{(1)}.p_1)(m_2+i\gamma^{(2)}.p_2) \Phi(q,P); \quad
 p_{1,2} = {\hat m}_{1,2} P \pm q.
\end{equation}
In terms of $\Phi$, eq.(3.1) in Gordon-reduced form reads as   
\begin{equation}\label{3.7}
\Delta_1 \Delta_2 \Phi(q,P) = -i(2\pi)^{-4} F_{12} \int d^4q' V_\mu^{(1)} 
V_\mu^{(2)} V({\hat q},{\hat q}') \Phi(P,q');
\end{equation}
where the $V_\mu$-functions are given by [22, 23, 44]
\begin{equation}\label{3.8}
V_\mu^{(1,2)}= \pm 2m_{1,2} \gamma_\mu^{(1,2)}; \quad 
V_\mu^{(i)}=p_{i\mu}+p_{i\mu}'+i \sigma_{\mu\nu}^{(i)}(p_{i\nu}-p_{i\nu}')    
\end{equation}
\par
	Now to implement the Transversality Condition [12] for the entire 
kernel of eq.(3.7), all time-like components $\sigma, \sigma'$ in the 
product $V^{(1)}.V^{(2)}$ must first be replaced by their $on-shell$ values. 
Substituting from (3.8) and simplifying gives [22, 44] 
\begin{equation}\label{3.9}
V^{(1)}.V^{(2)} = 4{\hat m}_1{\hat m}_2 P^2 -({\hat q}+{\hat q}')^2
-2({\hat m}_1-{\hat m}_2) P.(q+q') + ``spin-Terms'';
\end{equation}
\begin{equation}\label{3.10}
``Spin Terms'' = -i(2{\hat m}_1P+{\hat q}+{\hat q}')_\mu\sigma_{\mu\nu}^{(2)}
{\hat k}_\nu+i(2{\hat m}_2P-{\hat q}-{\hat q}')_\mu \sigma_{\mu\nu}^{(1)}
{\hat k}_\nu + 
\sigma_{\lambda\mu}^{(1)} \sigma_{\lambda\nu}^{(2)}k_\mu k_\nu
\end{equation}

\subsection*{3.2 Reconstruction of Fermion Vertex Fn} 

Eq.(3.7) is the fermionic counterpart of (2.3) for scalar quarks, with the 
common scalar function $\Phi$  linking the two descriptions, so that the 
3D reduction of (3.7) follows the  steps (2.3) to (2.4) with
the identification of $\phi$ as the appropriate 3D wave function. We skip the
counterpart of eq.(2.4) for brevity, except to note that it is the
appropriate dynamics for spectroscopy [22,31] under $CNPA$, with a formally 
identical algebraic structure as in $CIA$, so that the spectroscopic 
predictions of both must be the same. These spectroscopic details [44], 
are not of immediate concern here  except for the (gaussian) structure of $\phi$:
\begin{equation}\label{3.11}
\phi({\hat q}) = \exp{(-\frac{{\hat q}^2}{2\beta^2})} 
\end{equation} 
where the quantity $\beta^2$ is $dynamically$ determined in terms
of the input quantities (3.3-5) as [23, 39, 44]
\begin{eqnarray}\label{3.12}
\beta^4  &=& 2{\hat m}_1{\hat m}_2 M \omega_{q{\bar q}}^2/\gamma^2; \\ \nonumber
\gamma^2 &=& 1-\frac{2\omega_{q{\bar q}}^2 C_0}{M_> \omega_0^2}; \\ \nonumber
 M_>     &=&  \sup{(M, m_1+m_2)}
\end{eqnarray} 
and is of course a Lorentz invariant quantity (independent of $n_\mu$).
\par  
	With this knowledge of $\phi$, we may now reconstruct the 4D $fermion$ 
vertex function in two stages. First the auxiliary scalar $\Phi(q,P)$ satisfying eq.(3.7), is treated as in the steps (2.3-2.8) to express it terms of the 3D quantities $\phi$ and $D_n$, viz.,
\begin{equation}\label{3.13}
\Delta_1 \Delta_2 \Phi(q,P) = D_n({\hat q}) \frac {\phi({\hat q})}{2i\pi}
\end{equation} 	
Next the connection (3.6) between the the 4D fermionic $\Psi$ and the
auxiliary $\Phi$ function yields $\Psi$ directly in terms of the 4D
hadron quark vertex function:     
\begin{equation}\label{3.14}
\Psi(P,q)= S_F(p_1) \Gamma({\hat q}) \gamma_D S_F(-p_2); \quad
\Gamma({\hat q}) = N_H [P_n/M] D_n({\hat q})\frac {\phi({\hat q})}{2i\pi}
\end{equation}   
Here $\Gamma$ is a scalar factor carrying the bulk of the 
dynamical information, while $\gamma_D$ is a (kinematical) Dirac matrix 
which equals $\gamma_5$ for a P-meson, $i\gamma_\mu$ for a V-meson, 
$i\gamma_\mu \gamma_5$ for an A-meson, etc.
$N_H$ represents the hadron normalization which is formally defined by [23]:
\begin{equation}\label{3.15}
2iP_\mu = (2\pi)^4 Tr{ \int d^4 q [{\bar \Psi}i \gamma_\mu \Psi
(m_2-i\gamma.p_2){\hat m}_1 +(-{\hat m}_2){\bar \Psi (m_1+i\gamma.p_1)  
\Psi i\gamma_\mu}]}
\end{equation}
where the 4D wave function $\Psi$, together with its adjoint ${\bar \Psi}$,
are given by (3.14)and its adjoint equation respectively; ${\hat m}_i$,
given by eq.(3.4), are the Wightman-Gaerding definitions [45, 23] of the 
momentum fractions carried by the two quarks. The normalizer $N_H$, eq.(3.15),
is evaluated in Appendix C under both $CIA$ and $CNPA$ conditions. However,
if it were regarded as the zero momentum limit of the e.m. form factor of a 
hadron (via quark triangle loop), it would cause Lorentz mismatch problems [32] 
under $CIA$, which is a principal reason [39] for recourse to $CNPA$.

\subsection*{3.3  $CNPA$ Application to $\pi^0 \rightarrow 2\gamma$}  

We close this Section with the example of  two-photon
decay of a $\pi^0$ meson which is given by a triangle loop. Since  there
is one hadron-quark vertex in this case, it does not suffer from the 
Lorentz mismatch problem.The invariant amplitude for 
$\pi^0 \rightarrow \gamma \gamma$ decay under $CNPA$ conditions may be
written down from fig.1 below to give [11]

\begin{figure}[t]

\caption{}
\vspace{0.5in}

\begin{picture}(450,175)(-30,-10)
\put(35,10) {(a) $\pi^0 \Rightarrow \gamma_1\gamma_2$} 

\multiput(25,110)(-3,7){7}{\line(1,0){2.8}}
\multiput(25,110)(-3,7){7}{\line(0,1){7}}
\multiput(95,110)(3,7){7}{\line(-1,0){2.8}}
\multiput(95,110)(3,7){7}{\line(0,1){7}}

\put (27.8,110){\vector(1,0){33.6}}
\put (61.4,110){\line(1,0){33.6}}

\put(26,112){\makebox(0,0){\large O}}
\put(95,112){\makebox(0,0){\large O}}

\put(60,121){\makebox(0,0){$q-Q$}}

\put(60,90){\makebox(0,0){\large ${\nabla}$}}

\put(58.5,57){\rule{1mm}{10mm}}

\put(60,44){\makebox(0,0){$P$}}

\put(60,76){\line(-1,-2){5}}
\put(60,76){\line(1,-2){5}}

\put(37,94){\makebox(0,0){$p_1$}}
\put(10,170){\makebox(0,0){$\gamma(k_1)$}}

\put(83,94){\makebox(0,0){$p_2$}}
\put(112,170){\makebox(0,0){$\gamma(k_2)$}}

\put (28,110){\line(5,-3){28}}
\put (92,110){\line(-5,-3){28}}

\put (93.5,130){\line(3,2){10}}
\put (104,137){\line(1,-4){3}}

\put (16,137){\line(3,-2){10}}
\put (11,127){\line(1,2){5}}

\put(255,10) {(b) $\pi^0 \Rightarrow \gamma_2\gamma_1$} 

\multiput(245,110)(-3,7){7}{\line(1,0){2.8}}
\multiput(245,110)(-3,7){7}{\line(0,1){7}}
\multiput(315,110)(3,7){7}{\line(-1,0){2.8}}
\multiput(315,110)(3,7){7}{\line(0,1){7}}

\put (247.8,110){\vector(1,0){33.6}}
\put (281.4,110){\line(1,0){33.6}}

\put(246,112){\makebox(0,0){\large O}}
\put(315,112){\makebox(0,0){\large O}}

\put(280,121){\makebox(0,0){$q+Q$}}

\put(280,90){\makebox(0,0){\large ${\nabla}$}}

\put(278.5,57){\rule{1mm}{10mm}}

\put(280,44){\makebox(0,0){$P$}}

\put(280,76){\line(-1,-2){5}}
\put(280,76){\line(1,-2){5}}

\put(257,94){\makebox(0,0){$p_1$}}
\put(230,170){\makebox(0,0){$\gamma(k_2)$}}

\put(303,94){\makebox(0,0){$p_2$}}
\put(332,170){\makebox(0,0){$\gamma(k_1)$}}

\put (248,110){\line(5,-3){28}}
\put (312,110){\line(-5,-3){28}}

\put (313.5,130){\line(3,2){10}}
\put (324,137){\line(1,-4){3}}

\put (236,137){\line(3,-2){10}}
\put (231,127){\line(1,2){5}}

\end{picture}
\end{figure}

\begin{equation}\label{3.16}
A(\pi^0 2\gamma) = \frac{1}{\sqrt 6} e^2 Tr \int d^4 q [\Psi (q,P)
i\gamma.\epsilon^{(1)} S_F(q-Q) i\gamma.\epsilon^{(2)} + 1 \Rightarrow 2]
\end{equation}  
where $2q = p_1 - p_2$ and $2Q=k_1-k_2$, and 
the color and flavour factors have been taken in the standard way.
The second term corresponds to an interchange of the two photons. This
general structure defines the $\pi^0\gamma\gamma$ form factor $F_\pi$
through the relation [46]
\begin{equation}\label{3.17}
A(\pi^0 2\gamma) \equiv F_\pi \epsilon_{\mu\nu\rho\sigma}
\epsilon_\mu^{(1)} \epsilon_\nu^{(2)} P_\rho Q_\sigma
\end{equation}            		
Evaluating the traces in (3.16) after substitution from (3.14), and
a routine simplification, leads to the identification
\begin{equation}\label{3.18}
F_\pi = \frac{e^2}{\sqrt{6(2\pi)^3}} \frac{4m_q N_\pi}{2 i \pi} \int d^4 q
\frac{D_n \phi}{\Delta_1 \Delta_2}[ \frac{1}{\Delta_3^+} + 
\frac{1}{\Delta_3^-}]
\end{equation}
where
$$ \Delta_{1,2} = m_q^2 +p_{1,2}^2; \quad 
\Delta_3^{\pm} = m_q^2 + (q \mp Q)^2 $$
The gauge invariance is of course explicit from the structure of (3.17).
To simplify (3.18), the Lorentz invariant measure 
$$ d^4 q = d^2 q_\perp d(q.n)d(q_n)$$
may be used first to integrate over $d(q_n)$, to yield a remarkably
simple yet accurate result:
\begin{equation}\label{3.19}
\int d(q_n) \frac{D_n \phi}{\Delta_1 \Delta_2}[ \frac{1}{\Delta_3^+} + 
\frac{1}{\Delta_3^-}] \approx \frac{2i\pi \phi}{m_q^2 + q_\perp^2}
\end{equation}
which is a $CNPA$ adaptation of the corresponding result 
[47] on $\pi^0 2\gamma$ decay in usual null-plane variables [37].
(For details of steps on null-plane pole integrations in the $p_{2-}$ 
variable, see ref.[47]).  Eq.(3.18) then simplifies to
\begin{equation}\label{3.20}
F_\pi = \frac{4e^2m_q N_\pi}{\sqrt{6(2\pi)^3}} \int \frac
{d^2 q_\perp M dx \phi}{m_q^2 + q_\perp^2}         
\end{equation}  
where $\phi$ is given by eq.(3.11), with ${\hat q}^2 = q_\perp^2 + x^2 M^2$.
The integral finally works out at
\begin{equation}\label{3.21}
F_\pi = \frac{4e^2m_q N_\pi \beta^3}{\sqrt{6}} 
erf(\sqrt{M^2/8\beta^2}) \int_0^\infty dx \frac{e^{-x}}{m_q^2+2\beta^2 x}    
\end{equation}
The decay rate in turn is given by [46]
\begin{equation}\label{3.22}
\Gamma(\pi^0 \rightarrow 2\gamma) = \frac{F_\pi^2 M^3}{64 \pi}
\end{equation} 
Using the value of $N_\pi = 31.88 GeV^{-3}$ after substitution from
eq.(C.5) of Appendix C, $F_\pi$ is predicted as $29 MeV^{-1}$, leading
to the value $11 ev$ which agrees with the $CIA$ value [11] but is 
about $30 \%$ higher than the observed value of $8.5 ev$ [13]. We note
in passing that an alternative formulation in terms of ``half-off-shell''
wave functions in null-plane variables [23] gives a much closer agreement
with experiment [13]. However such wave functions [23], although conforming
to the Weinberg [6] spirit of the infinite momentum frame, fail to
satisfy the angular condition [35, 9] necessitated by $O(3)$ invariance.

\section*{4  Dynamical Mass Via $DB{\chi}S$  Scenario}
 
The `dynamical' mass function of the quark  may be defined in 
one of two ways: i) as the non-trivial solution of the SDE [27] 
under $DB{\chi}S$ [28]; ii) as the vertex function $\Gamma({\hat q})$, 
for the pion in the chiral limit $(M_\pi^2 = 0)$. The logic of the second form  
follows from the original NJL paper [28] for contact interaction,
which was subsequently found to be more generally satisfied for
extended 4-fermion interaction with vector exchange [27] whose 
chiral invariance ensures that the SDE for the self-energy operator
$\Sigma(p)$ (essentially the quark mass function $m(p)$), and the 
BSE for the  pion-quark vertex function $\Gamma(q,P)$ are formally 
$identical$ in the limit of zero pion 4-momentum, leading to the
conclusion that these two functions are basically the same,
except for the normalization. This result is also valid for the 
$MYTP$ oriented 3D-4D BSE formalism [30], except for the replacement 
of $m(p)$ by $m({\hat p})$, and offers a practical way to construct
the mass function in terms of the pion-quark vertex function via the
BSE route for hadron-quark interaction [30]. 

\subsection*{4.1  Mass Fn as $DB{\chi}S$ Limit of Pion Vertex Fn}
\setcounter{equation}{0}
\renewcommand{\theequation}{4.\arabic{equation}}                  

The general hadron-quark vertex function is proportional to the 
product $D({\hat q}) \times \phi({\hat q})$, so that the mass
function $m({\hat p})$ is obtained by setting $M_\pi=0$ in this 
expression in the limit $P_\mu=0$, where $p_\mu$ is now the 4-momentum 
of either quark. Making the necessary substitutions, the mass function 
is identified in  $CIA$  as [30]:
\begin{equation}\label{4.1}
m({\hat p}) = \frac{\omega^3({\hat p})}{m_q^2} \phi({\hat p}); \quad
\omega^2({\hat q}) = m_q^2 + {\hat p}^2,
\end{equation}
normalized to the `constituent' mass $m_q$ in the limit of ${\hat p}=0$.    
As a simple check, the mass function vanishes in the $p\rightarrow \infty$
limit. The 3D wave function  $\phi$ has the gaussian form (3.11), with 
$\beta^2 = 0.060 GeV^2$ [39] after substitution from eq(3.12), together
with (3.4-5), for the pion case.   
\par
	We now turn to the corresponding derivation under $CNPA$
in close parallel to above [30], except for the definition of the
denominator function which we write in the `standard' null-plane 
notation [37-38,23] for easier comparison with $CIA$ :
\begin{equation}\label{4.2}
D_n = 2P_+ (m_q^2 + {\hat q}^2- M^2/4); \quad 
{\hat q}^2 = q_\perp^2 + M^2 q_+^2/P_+^2 
\end{equation} 
The factor in front shows that the role of $2\omega({\hat q})$ in 
the instant form ($CIA$) is now played by $P_+$ in the null-plane form. 
This is in conformity with the Dirac-Weinberg notion [34,6] of the
`plus' component as the `mass' term, which is of course 
orientation ($n_\mu$)-dependent. The $CNPA$ mass function is now
\begin{equation}\label{4.3}
m_+({\hat p}) = p_+ \frac{(m_q^2+{\hat p}^2) \phi({\hat p})}{m_q^2}; 
\quad p_+ = {\sqrt 2}p.n 
\end{equation}
in the same relative normalization as in eq.(3.1), and with the 
replacement $P_+(=p_{1+} + p_{2+}) \Rightarrow 2p_+$ in the chiral limit.                
\par
	This form of the mass function is convenient for applications
to certain types of loop integrals such as vacuum condensates [30]
among other things. It is not of course Lorentz invariant by itself, 
unlike in standard 4D SDE-BSE formalism [27], but this is not
a serious problem since it is not a directly measurable quantity
except in the limits of $p \rightarrow 0$ (constituent mass), or
$p \rightarrow \infty$ (current mass), where it is Lorentz invariant. 
However it yields Lorentz invariant quantities where it enters as
a dynamical ingredient, e.g., in the evaluation of vacuum 
condensates [30]; see Section 5.     

\subsection*{4.2  Dynamical Mass from SDE for $\Sigma (p)$ }
 
The more standard aspect of the `dynamical' mass function is its
appearance as the non-trivial solution of the SDE under $DB{\chi}S$ [27]. 
We now give a summary derivation of the 3D-4D counterpart of this basic 
result, which although obtained under $CIA$  premises, is almost 
literally valid for  $CNPA$, with the replacement $k_l \rightarrow k_n$.  
To that end we start with the non-perturbative part of the gluon 
propagator $D_{\mu\nu}(k)$ = $D(k)[\delta_{\mu\nu}-k_\mu k_\nu/k^2]$ for 
the (harmonic) interaction of $ud$ quarks  
where the scalar factor $D(k)$ has the form [30]
\begin{equation}\label{4.4}
D(k)= {3 \over 4}(2\pi)^3 \omega_0^2 2m_q {\alpha_s(4m_q^2)}
[\nabla_{\hat k}^2 +C_0/\omega_0^2] {\delta^3({\hat k})}
\end{equation}
which is immediately derivable from the structure of the `potential'
function $V{\hat k})$, eq(3.2), with the $A_0$-term dropped as 
insignificant for this case, and taking $M_>=2m_q$ for the `pion'. 
Note that $D({\hat k})$ has a directional dependence $n_\mu=P_\mu/\sqrt{P^2}$ 
on the pion 4-momentum $P_\mu$, so that  ${\hat k}^2 >0$ over all 4D 
space; it also possesses a well-defined limit for $P_\mu \rightarrow 0$. 
This structure may now be substituted in the SDE for a self-consistent 
solution in the low momentum limit, which in the Landau gauge 
$A(p^2)=1$ [48] becomes [30]
\begin{equation}\label{4.5}
m(p)= \frac{3i}{\pi} \int d^3{\hat k}dk_0 m_q \alpha_s 
[\omega_0^2 {\nabla_{\hat k}}^2 +C_0] \delta^3({\hat k}) 
\frac{m(p'^2)}{[p'2+m^2(p'^2)]}
\end{equation}
where $p'=p-k$ is 4D, and (${\hat k}$, $k_0$) are (3D,1D) respectively.
The integration is essentially over the time-like $k_0$, with the `pole' 
position at ${p_0}'=m(p_0') \equiv m_{NJL}$, leading finally to [30]
\begin{equation}\label{4.6}
m_{NJL}={{3m_q \alpha_s} \over {m_{NJL}}^2} [3\omega_0^2 - C_0 m_{NJL}^2];
\quad \alpha_s = \frac{6\pi/29} {ln(10 m_q)}
\end{equation}
after substituting the values (3.4-5) for the QCD constant $\Lambda$, etc.   
The further identification of $m_q$ with $m_{NJL}$ in this equation, yields
an independent self-consistent estimate $m_{NJL} \sim 300 MeV$, which may be 
compared to the input value $265 MeV$, eq.(3.5) used for the spectra [31].
Thus the use of the SDE in conjunction with the BSE provides a powerful
check on the consistency of the otherwise empirical constituent mass which
is no longer a free parameter. This analysis so far ignores the Politzer 
relation [29] $m_{ud}=m_c+m_{NJL}$, for the constituent mass $m_q$ away 
from the chiral limit; for this extended derivation see [30].   
\par
	We end this Section with some comments on the interpretation of 
the two basic constants $C_0$ and $\omega_0$, in view of their appearance 
in the determining equation (4.6) for the constituent mass $m_{NJL}$. From
eq.(3.3), $\omega_0$ may be regarded as a `reduced spring constant' of 
the confining interaction for $light$ quarks (for which the constant $A_0$
is not important). It controls the confinement scale [30] for a hopefully
integrated view of the different flavour sectors of hadron spectra [31].
$C_0$ is a second constant designed to simulate the zero-point (vacuum)
energy effects via the replacement $r^2\rightarrow r^2-C_0/\omega_0^2$.
Both these quantities are as fundamental in a `potential' model context
[17],as the pionic constant $f_\pi$ is in, say, chiral perturbation theory 
[49], or the role that vacuum condensates play as cofficients of the
successive `twist' terms in the Wilson $OPE$ expansion employed in 
QCD sum rules [50]. In the present state of the QCD art, it is perhaps
a matter of taste as to which set of constants should be considered as 
more basic than the other, but the facility of a derivation of the latter
in terms of the former, as partly illustrated in the foregoing, 
should hopefully constitute a connection between the two languages, with 
the advantage of the `spectroscopic link' associated with the former [30]. 
The formal possibility of a self-consistent derivation of $m_{NJL}$ in
terms of $\omega_0$ and $C_0$, as illustrated above (while leaving scope
for quantitative corrections due to the neglected effects such as the $oge$ 
term), is one such manifestation of this connection.

\section*{5  $Direct$ Calculation Of Quark Condensates}

\setcounter{equation}{0}
\renewcommand{\theequation}{5.\arabic{equation}}                  

As was first shown by the Orsay group [51],  the `potential' method
offers a $direct$ method of calculation of the condensate, in terms 
of the quark's {\it non-perturbative} mass function $m(p)$ as the 
chiral ($M_\pi =0$) limit of the pion-quark vertex function 
$\Gamma({\hat q})$, viz., eq.(4.1) for $CIA$ or (4.3) for $CNPA$. 
This function must be used in the expression for the full propagator, 
$S_F(p)$ which appears in the formal definition of the condensate 
as follows: 
\begin{equation}\label{5.1} 
<{\bar q}q> =  {{iN_c N_f} \over {(2\pi)^4}} Tr {[\int d^4 p S_F(p)]}
\end{equation}
where 
$$ S_F(p) = \frac{-i}{m(p) + i\gamma.p} $$
in the Landau gauge [48]. Here $N_c=3$, and $N_f=1$ (since each 
separate flavour $(u/d)$ is counted). In the $MYTP$ scenario, the
mass function does $not$ depend on the time-like component of $p_\mu$.
Therefore after taking the traces on the RHS of (5.1), and  doing 
the pole-integration over the time-like component of $p_\mu$, the
above equation becomes for $CIA$ 
\begin{equation}\label{5.2}
<{\bar q}q> = - \frac{3}{4{\pi}^3} \int d^3{\hat p}\frac {m({\hat p})}
{\sqrt {{\hat p}^2 + m^2({\hat p})}}
\end{equation}
To evaluate the 3D integral (5.2) further, substitute the $CIA$ 
structure (4.1) for $m({\hat p})$, with $\phi({\hat p})$=
$exp(-{\hat p}^2/{2\beta^2})$, which gives a simple quadrature for the
resulting integral. Further, since the integral has an analytic form
in $m_q$, it is useful for evaluating a related quantity, viz., the
`increment' ${\delta <{\bar q}q>}$ due to a shift ${\delta m_q}$ in
the `constituent' mass, which by Politzer Additivity [29] equals a
corresponding shift ${\delta m_c}$ in the `current' mass. Both these
parameters are directly comparable with corresponding estimates from
QCD sum rules [50]. Using the inputs from (3.4-5) gives 
$\beta^2$ = $0.0603$, and the final results for this case are [52] 
\begin{equation}\label{5.3}
<{\bar q}q> = -(266 MeV)^3 ; \quad {\delta <{\bar q}q>} 
= + 0.0664 {\delta m_c}
\end{equation} 
These values are fully rooted in spectroscopy but are otherwise free from 
adjustable parameters, except for the quantity ${\delta m_c}$ which
represents the $u-d$ mass difference. The  
condensate has a fair overlap with QCD-SR determinations [48], but 
its increment is rather small (for possible reasons, see below).   
\par
	In a similar way the corresponding condensate results under $CNPA$
are found by substituting (4.3) in (5.1), to give
\begin{equation}\label{5.4} 
<{\bar q}q> =  \frac{12i{\sqrt 2}}{(2\pi)^4} \int d^3 {\hat p} dp_n 
\frac{p.n [1+{{\hat p}^2 \over {m_q^2}}] \phi({\hat p})} 
{m_q^2 +p_\perp^2 -2 p.n p_n}
\end{equation} 
The integration over $p_n$ is again trivial and the $CNPA$ counterpart
of (5.2) is :
\begin{equation}\label{5.5}
<{\bar q}q> ={{-3{\sqrt 2}} \over {(2\pi)^3}} \int d^3{\hat p}
[1+{{\hat p}^2 \over {m_q^2}}] \phi({\hat p}) 
\end{equation}
Substituting the gaussian form for $\phi$ and integrating, yields a 
simple analytic form: 
\begin{equation}\label{5.6}
<{\bar q}q> = -3{\sqrt 2}(\beta^2/{2\pi})^{3/2}[1+3\beta^2/m_q^2] 
= - (242 MeV)^3 
\end{equation}  
a value which seems to be even closer to the estimate $-(240)^3$ of 
QCD-SR [48] than the $CIA$ result $-(266)^3$  [52].  
\par 
	We end this Section by noting that the quantity 
${\delta <{\bar q}q>}$ offers a  comparison with QCD-SR [50] in terms of 
its effect on certain physical quantities derivable from it. Thus it 
contributes to hadron mass splittings due to strong SU(2) breaking [52],
albeit by a small amount. This contrasts with the corresponding 
QCD-SR findings [53] that suggest dominance of this very contribution. 
This is not surprising since within a BSE-cum-SDE framework, most of the  
non-perturbative effects are already contained in the hadron-quark vertex 
function,  with a correspondingly smaller role for the  condensates. 
This philosophy of the BSE-SDE formalism  is somewhat akin to that 
of the Pagels-Stokar [54] ``Dynamical Perturbation Theory'' (neglect of 
`criss-cross' gluon lines in a loop diagram),  which must be carefully 
distinguished from a naive interpration of  perturbative QCD.  
On the other hand in a QCD-SR scenario [50] such condensate
contributions  which arise  from the `twist terms' in an OPE expansion,
are perhaps the dominant source of non-perturbative effects.   

\section*{6  Resume And Conclusions}

Salpeter-like equations [1] which may be defined as BSE's with
3D kernel support, have the property of exact 3D reduction, as well as 
reconstruction of the 4D amplitude in terms of 3D ingredients. The
fact that such equations are governed by a well-defined dynamical principle
(with a gauge content [20]), known as the the  Markov-Yukawa 
Transversality Principle ($MYTP$) [12], gives them a unique status 
in the contemporary literature, characterized as they are by a two-tier 
dynamics, the 3D form for $O(3)$-like spectra, and the 4D form for 
transition amplitudes as 4D loop integrals. However the vehicle of 
$CIA$ through which $MYTP$ has operated so far, suffers from a sort of``Lorentz incompatibility'' [39] among the participating vertices in triangle loops and above. These show up as ill-defined integrals due to the presence of  time-like momenta 
in the (gaussian) form factors, leading to complexities in amplitudes [32],
while loops involving up to two quark lines [33] just escape this pathology.
To deal with this problem, we have proposed a generalization  from $CIA$ to a 
covariantly defined null-plane ansatz ($CNPA$) which retains the property
of 3D-4D interlinkage, but does $not$ suffer from  the problem of time-like 
gaussians in the loop integrals, albeit  at the cost of dependence
on the orientation $n_\mu$ of the null-plane [39]. However, as found from
a recent calculation of the pion form factor [39],  the
$n$-dependence is a mere technicality which may be trivially
eliminated via the `Lorentz completion' trick  leading to an
explicitly Lorentz-invariant structure [39]. In this paper an attempt has been 
made for a systematic development of the $CNPA$ framework on closely parallel
lines to $CIA$ through a few basic calibrations (the quark mass 
function, quark condensate and electroweak constants). Comparison 
with the $CIA$ framework is further facilitated by the fact that the 
reduced 3D forms have formally identical structures for both. 
However the predictions differ at the level of loop integrals: 
The difference is small for two-quark loops, but only $CNPA$ seems to make 
sense for triangle (and higher) loops [39], for which $CIA$ is ill-defined.   

\subsection*{6.1  What Distinguishes Salpeter-like Eqs From Others ?}    

We conclude with a  summary of some salient features of Salpeter-like Eqs [1]
which distinguish them from most other 3D approaches to strong interaction 
dynamics [1-3], including null-plane dynamics [6,9,35]. 
\par
	A first comparison of Salpeter-like Eqs. with a standard 4D BSE [27, 55]
concerns the role of some additional length scales in the infra-red part 
of the gluon propagator [55] which are $not$ quite `tested' without
considering  $L-excited$ spectra, for which there is no evidence yet [55].  
$MYTP$-based Salpeter-like Eqs, on the other hand, allow a more explicit test
of the gluon propagator [31] in respect of both ground and $L$-excited states.  
\par
	Another aspect concerns the question of the 3D support ansatz 
going beyond the conventional ladder approximation, since the very interpretation
of the quantity $m_{NJL}$ as a `constituent mass' could otherwise be 
questioned on the ground  that its generation requires the presence of a 
$second$ $source$ of color charge [56], while the solution of the SDE  in 
the rainbow approximation [55] misses this detail due to the dependence of the standard $oge$ propagator on a $single$ 4-vector $k_\mu$ only. On the other hand, 
a Salpeter-like Eq, via the 3D support ansatz, effectively ensures that the $oge$
propagator ``sees'' [30] the second source through its directional dependence
on the composite 4-momentum $P_\mu$, in addition to $k_\mu$. Indeed 
the identification of the mass function $m(p)$ as the chiral limit
($P_\mu \rightarrow 0$) of the pion-quark vertex function $\Gamma(q,P)$
(which in turn is a solution of a 2-body equation), would not be 
consistent if $m(p)$ were to depend on $p_\mu$ alone. A Salpeter-like Eq,
with its $MYTP$ based dependence on 
$${\hat p}_\mu = p_\mu - p.P P_\mu/P^2 $$ 
makes this function logically more consistent with the concept of a
second source of color charge [30, 56].               
\par
	A last item of comparison concerns the elimination of a class of 
singularities which would appear in a 4D loop integral due to overlapping 
pole effects, viz., the Landau-Cutkowsky singularities which are usually 
responsible for `free' propagation of quarks inside closed loops. According 
to the standard wisdom [57], the infrared behaviour of the gluon propagator 
helps tone down the effect of this pole, via WT identities, but doubts have 
also been expressed [57] about the uniqueness of the procedure. In an  
($MYTP$-governed) Salpeter-like Eq, the characteristic $D \times \phi$ 
structure of the hadron-quark vertex function  automatically 
ensures that the $D$-function simply cancels out such Landau-Cutkowsky 
poles, and thus prevents the free propagation of quarks. This is true for
both $CIA$ [11] and $CNPA$ [39] by virtue of the $D\times \phi$ form of
the vertex function.       
\par
	The 3D-4D interlinkage offered by $MYTP$ is also generalizable to a
3-body BSE with pairwise kernels under covariant 3D support [26]. Other
applications include 3-hadron couplings like $\rho-\pi-\pi$, 
$\omega-\rho-\pi$, tests of Vector Meson Dominance, etc., some of 
which are under way.    
\par
	One of us (BMS) is grateful to Prof R.K.Shivpuri for the hospitality
of the High Energy Lab at Delhi University.

\section*{Appendix A.  $MYTP$ As A Gauge Principle}

\setcounter{equation}{0}
\renewcommand{\theequation}{A.\arabic{equation}}                  

In this Appendix we summarise the gauge aspects of $MYTP$ [12b] 
which are brought out by the dynamics of bilocal fields [19] and
demonstrate the 3D-4D interlinkage of BS amplitudes [11].   
Now the gauge principle underlying $MYTP$ asserts the redundance [20]
of the relative `time' variable $x_0$, ($x=x_1-x_2$), whose covariant
definition is just the longitudinal component of $x_\mu$ in the direction 
of $P_\mu$, viz.,  $x.P P_\mu/P^2$. This `redundance' is expressed 
by the statement that a translation of the relative coordinate [20]
$x_\mu \rightarrow x_\mu' + \xi P_\mu$ on the bilocal field ${\cal M}(x,P)$:
$${\cal M}(x_\mu,P_\mu) \rightarrow {\cal M}_\xi(x_\mu,P_\mu)
= {\cal M}(x_\mu+\xi P_\mu,P_\mu)$$,
which is a  sort of `gauge transformation' for the bilocal field [20], 
should leave this quantity $invariant$. This invariance is just 
the content of the Markov-Yukawa subsidiary condition [12] which, 
under an interchange of the relative coordinates and the momenta reads 
as [20, 5b]
\begin{equation}\label{A.1}
P_\mu {{\partial} \over {\partial x_\mu}} {\cal M}(x_\mu,P_\mu)=0 
\end{equation} 
where the direction $P_\mu$ guarantees an irreducible representation of the 
Poincare' group for the bilocal field ${\cal M}$ [20]. An equation of this
type has been used in other approaches to bilocal field dynamics (see
ref [20] for other references), but this `gauge' interpretation of the
subsidiary condition [20] provides a more transparent view of the same
condition which we have abbreviated as $MYTP$ above. 
\par
	Eq.(A.1) amounts to an effective 3D support to the interaction
between the constituents of the bilocal field, which may be alternatively
postulated directly for the pairwise BSE kernel $K$  by demanding that 
it be a function of only ${\hat q}_\mu = q-q.PP_\mu/P^2$, which implies that 
${\hat q}.P \equiv 0$. This approach in which the propagators retain their 
full 4D forms, allows the use of both the 3D and 4D BSE forms in an
interchangeable manner, as shown below. 
 
\subsection*{A.1  3D-4D Interconnection: Spinless Particles} 

To demonstrate the basic 3D-4D interconnection  under $MYTP$ [12], consider  
a system of two identical spinless particles, with the BSE [11]  
\begin{equation}\label{A.2}
 i(2\pi)^4 \Phi(q,P) = (\Delta_1\Delta_2)^{-1} \int d^3{\hat q}'M d\sigma' 
K({\hat q},{\hat q}') \Phi(q',P);  [\Delta_{1,2} = m_q^2 +p_{1,2}^2]
\end{equation}
where the 3D support to the kernel $K$ is implied in its `hatted' structure:
\begin{equation}\label{A.3}
{\hat q}_\mu= q_\mu- \sigma P_\mu; \sigma= q.P/P^2; {\hat q}.P \equiv 0.
\end{equation}                                                                            
The relative and total 4-momenta are related by

$$p_1+p_2 = P = p_1'+p_2'; 2q = p_1-p_2; 2q'=p_i'-p_2'.$$     

The 3D wave function $\phi({\hat q})$ is defined by [12]
\begin{equation}\label{A.4}
 \phi({\hat q}) = \int M d\sigma \Phi(q,P) 
\end{equation}
When (A.4) is substituted on the RHS of (A.2) one gets
\begin{equation}\label{A.5}
 i(2\pi)^4 \Phi(q,P) = (\Delta_1\Delta_2)^{-1} \int d^3{\hat q}' 
K({\hat q},{\hat q}') \phi({\hat q}')
\end{equation}

Now integrate both sides of this equation wrt $\sigma$ to get an explicit
3D equation
\begin{equation}\label{A.6}
 (2\pi)^3 D({\hat q}) \phi({\hat q}) = 
\int d^3{\hat q}' K({\hat q},{\hat q}') \phi({\hat q}')           
\end{equation}
where the 3D denominator function is given by
\begin{equation}\label{A.7}
 2i\pi D^{-1}({\hat q}) = \int M d\sigma (\Delta_1 \Delta_2)^{-1} 
\end{equation}
A comparison of (A.5) with (A.6) via (A.7)  gives the 3D-4D interconnection  
\begin{equation}\label{A.8}
 2i\pi \Delta_1 \Delta_2 \Phi (q,P) = D({\hat q}) \phi({\hat q})
\end{equation}
which directly identifies the RHS as the hadron-quark Vertex Function 
\begin{equation}\label{A.9}
\Gamma = D \times \phi /2i\pi.                                   
\end{equation}

\section*{Appendix B   Salpeter Equation: 3D-4D Interlinkage}

\setcounter{equation}{0}
\renewcommand{\theequation}{B.\arabic{equation}}                  

In this Appendix we sketch the main steps [15] to demonstrate the
3D-4D interlinkage of the BS amplitudes which stem from the Salpeter 
equation [1] for the relativistic hydrogen atom problem, in the 
notation of the original paper [1] : 
\begin{equation}\label{B.1} 
i\pi^2 F(q_\mu) \psi(q) = \alpha \int d^4 k {\bf k}^{-2} \psi(q+k) 
\end{equation}          
A comparison of this equation with eq.(A.6) shows a precise correspondence,
except for certain technicalities arising from its fermionic content.Indeed,
the 3D nature of the kernel in (B.1) is seen from its dependence on the 
3-vector ${\bf k}$, while the quantity $F(q_\mu)$ plays just the role of 
the product of the two 4D propagators $\Delta_1$ and $\Delta_2$ in (A.2): 
due to the (non-covariant) instantaneous (adiabatic) assumption [1]. 
\begin{equation}\label{B.2}
 F(q) = (\mu_1 E -H_1({\bf q})+\epsilon)(\mu_2 E -H_1({\bf q})-\epsilon)  
\end{equation}
with the time-like components identified as the $\epsilon$ terms !
Next, define the 3D wave function $\phi({\bf q})$ by
\begin{equation}\label{B.3}
 \phi({\bf q}) = \int d\epsilon \psi ({\bf q}, \epsilon)
\end{equation}          
which is the counterpart of (A.4), and use this result to integrate both
sides of (B.1) wrt $\epsilon$, after dividing by $F(q)$, so as to get
the 3D Salpeter equation [1]
\begin{equation}\label{B.4}
 [E-H_1({\bf q})-H_2({\bf q})]\phi_{\pm \pm} = 
\pm \Lambda_{\pm}^{(1)}\Lambda_{\pm}^{(2)} (-2i\pi \Gamma({\bf q})
= (-4i\alpha) \int d^3 k {\bf k}^{_2} \phi({\bf q+k})
\end{equation}
where the $\pm$ components are associated with the energy projection 
operators $\Lambda$ which however do not involve the time-like $\epsilon$.

The crucial aspect, on the other hand, is the 3D-4D interconnection  
which is obtained by substituting the second part of 
eq.(B.4) on the RHS of (B.1), after making use of (B.3):
\begin{equation}\label{B.5}
 F(q)\psi(q) = \Gamma({\bf q}) 
\end{equation}
where $\Gamma({\bf q})$ is the 3D BS vertex function. It is the precise
fermionic counterpart of the scalar eq.(A.9), since the $F(q)$ function is the 
product of the two 4D propagators. The form (B.5) is not formally covariant, 
but this is a mere technicality which can be remedied by standard methods. 
The reconstructability of the 4D vertex function from the Salpeter equation was independently noticed by the Bonn group [24]. The $MYTP$ [12] now provides a 
formal theoretical basis for all Salpeter-like equations in a $two-tier$ form [21-24]. 

\section*{Appendix C: BS Normalizers $N_P$ And $N_V$}

\setcounter{equation}{0}
\renewcommand{\theequation}{C.\arabic{equation}}                  

In this Appendix we outline a derivation of the BS normalizer $N_H$
from its `classical' definition [58, 59]  
\begin{equation}\label{C.1}
\frac{2iP_\mu}{(2\pi)^4}= \int d^4q Tr [{\bar \Psi(q,P)}
\partial_{P_\mu} S_F(-p_2)^{-1} \Psi(q,P)S_F(-p_2)^{-1}]
+ (1 \Leftrightarrow 2)
\end{equation} 
which is of course fully equivalent to (3.15) of text.  To illustrate
the techniques of both the $MYTP$ scenarios, we consider first 
the pseudoscalar case under $CIA$, followed by the vector case
under $CNPA$.

\subsection*{C.1 \quad  Pseudoscalar Meson Normalizer}

To simplify (C.1), note that

$$ \partial_{P_\mu}S_F^{-1}(\pm p_{1,2}) = \pm i \gamma_\mu {\hat m}_{1,2}
+ i\gamma_\mu \sigma $$

where $\sigma \equiv q.P/P^2$ is the longitudinal fraction of $q$ in the
$P$-direction. Substituting in (C.1) from eq.(3.14) of text, and taking
the traces, the result for pseudoscalar mesons is 
$$ N_P^{-2}= 2\int \frac{d^4q}{i(2\pi)^4}\frac{D^2\phi^2}
{\Delta_1^2 \Delta_2} [(M^2-{\delta m}^2 +\Delta_2)({\hat m}_1+\sigma)^2   
+ \Delta_1 ({\hat m}_1+\sigma)] + (1 \Leftrightarrow 2)$$     
where ${\delta m}= m_1-m_2$. Unforfunately the $({\hat m}_1+\sigma)^2$
term in the numerator causes a $negative$ contribution to the 
$\sigma$-integration. To overcome this problem, one may consider, following 
Nishijima [60], that the `charge' is concentrated on $one$ of the quark
lines, say $p_1$, which amounts to taking the derivative w.r.t. $p_1$
instead of w.r.t. $P$ as above. A more symmetrical possibility consists 
in interchanging $p_1$ with $p_2$, and weighting these two contributions
with the momentum fractions ${\hat m}_{1,2}$ respectively. The result
for the pseudoscalar case, after simplification is 
\begin{eqnarray}\label{C.2}         
N_P^{-2} &=& 2{\hat m}_1 \int \frac{d^4q}{i(2\pi)^4}\frac{D^2\phi^2}
{\Delta_1^2 \Delta_2} [(M^2-{\delta m}^2 +\Delta_2)({\hat m}_1+\sigma) \\  \nonumber
         & & + \Delta_1] + (1 \Leftrightarrow 2)
\end{eqnarray}
So far the treatment is quite general.We now specialize to the $MYTP$ [12] 
derivation under $CIA$ conditions [11]. Since the normalizer corresponds to
an e.m. form factor (with $k_\mu = 0$), it just escapes the problem of time-like components in the gaussian form factors, even in the $CIA$ scenario [33].   
\par
	The `pole' integration over $M d\sigma$ may now be carried out 
as in Sect.2 [23], and the result is a 3D integral:
\begin{equation}\label{C.3}
N_P^{-2}= \int \frac{d^3{\hat q}\phi^2({\hat q})}{(2\pi)^3} [(1-\frac
{{\delta m}^2}{M^2}) G({\hat q})+ (1+\frac{{\delta m}^2}{M^2})D({\hat q})
\end{equation}
where
$$G({\hat q})= 2 (\omega_1\omega_2+M^2{\hat m}_1{\hat m}_2)\omega_{12}$$
and the other symbols are as defined in Sect.2.  The rest of the 
quadrature is routine, but is skipped for brevity.  Since the $CNPA$
techniques are illustrated below (for V-mesons), we give without proof
the corresponding Lorentz-invariant result under $CNPA$ [39], viz.,   
\begin{equation}\label{C.4}
N_P^{-2}=2M \int \frac{d^3{\hat q}}{(2\pi)^3}
e^{-{\hat q}^2/\beta^2} [(1+{\delta m}^2/M^2)({\hat q}^2-\lambda/{4M^2})
+2{\hat m}_1{\hat m}_2 (M^2-{\delta m}^2)]           
\end{equation}
where the internal momentum ${\hat q}=(q_\perp,Mx)$ is formally a 3-vector,
in conformity with the `angular condition' [9,23,35]. 
Specializing to equal mass kinematics (needed for $F_\pi$ calculation 
in sect 3.3), we have
\begin{equation}\label{C.5}
N_P^{-2} = 2M \int_{-1/2}^{+1/2}M \frac{dx \pi \beta^2}{(2\pi)^3} 
e^{-M^2 x^2/\beta^2}[\beta^2 + M^2 x^2 + m_q^2 + M^2/4]  
\end{equation}
which works out at
$$ N_P^2 = 2M \frac{\beta^3}{(\pi)^{3/2}} [-\frac{M\beta}{2\sqrt{\pi}}
e^{-M^2/{4\beta^2}} +(3\beta^2/2 +m_q^2 +M^2/4) erf{(M/2\beta)}] $$
   
\subsection*{C.2 \quad Vector Meson Normalizer}

We next consider the Vector meson normalization directly under the 
Covariant null plane ($CNPA$) formalism of Sect.2, and under  
unequal mass kinematics.         
\par
	The BS normalizer $N_V$ for a V-meson is obtained from
(C.1) with the replacement $\gamma_5 \rightarrow i\gamma.\epsilon$,
before evaluating the traces. The V-meson result analogous to (C.2) is
\begin{eqnarray}\label{C.6}
N_V^{-2} &=& 2{\hat m}_1 \int \frac{d^4q}{i(2\pi)^4}\frac{D^2\phi^2}
{\Delta_1^2 \Delta_2} [\Delta_1 \\  \nonumber
         & & + (M^2-{\delta m}^2 +2m_1^2/3 + 2m_2^2/3 
-2\Delta_1/3)({\hat m}_1-\Delta_1/{2M^2})] \\  \nonumber
         & &  + [1 \Leftrightarrow 2]
\end{eqnarray}
where we have dropped all the $\Delta_2$ terms in the first group, 
anticipating their vanishing on pole integration [23]. We have also 
averaged over the V-meson polarization according to
$$ < \epsilon_\mu\epsilon_\nu > = \frac{1}{3}\theta_{\mu\nu}; \\
\theta_{\mu\nu}\equiv \delta_{\mu\nu}+\frac{P_\mu P_\nu}{M^2} $$
The last term on the RHS of (C.6) has been simplified from the term
$$ 4\epsilon.p_1 \epsilon.p_2 p_{1\mu}-2 \Delta_1 \epsilon.p_2 \epsilon_\mu$$
in the trace calculation, in the same relative normalization as the others
(after extracting an overall factor $P_\mu$), and is explained as follows. 
Since the only component that survives after integration is in the 
direction of $P_\mu$,  the second term (proportional to 
$\epsilon_\mu$) vanishes due to $\epsilon.P=0$. For the same reason,
we have symmetrically
$$4p_1.\epsilon p_2.\epsilon = -2(p_1.\epsilon)^2 -2(p_1.\epsilon)^2 $$
which on averaging over the vector polarization gives  
$$ +2(m_1^2+m_2^2)/3 - 2(\Delta_1+\Delta_2)/3
- [2(p_1.P)^2+2(p_2.P)^2]/{3M^2} $$
In the last term which has a second order of virtuality (in powers
of $\Delta_i$), we invoke the $CNPA$, viz., $p_- = -p_+ M^2/P_+^2$ for 
each of $p_{1,2}$, so that 
$$(p_i.P)^2 =[(p_{i-}P_+ +p_{i+}P_-)/2]^2 = 0; \quad P_- = M^2/P_+; $$ 
[However we do $not$ take this liberty in the 
linear terms in $p_i.P$ which have only a first order of virtuality].
Thus we take $ p_{1\mu} = -p_1.P P_\mu/M^2$ with
$$ p_1.P = (\Delta_1-\Delta_2)/2 -M^2{\hat m}_1; \quad 
2M^2{\hat m}_1 = M^2+m_1^2-m_2^2 $$    
\par
	For integration over $d^4q$, it is convenient to convert to
new variables as follows [37, 23]
$$ d^4 q = d^2 q_\perp dp_{2+}dp_{2-}/2 $$ 
where the pole integration is first carried out over $p_{2-}$ to give:
$$ \int dp_{2-}\frac{D^2}{4i\pi \Delta_1^2 \Delta_2} = 2p_{2+}M^2/P_+^2 $$
$$ \int dp_{2-}\frac{D}{4i\pi \Delta_1 \Delta_2} = M/P_+ $$ 
\begin{eqnarray}\label{C.7} 
N_V^{-2} &=& 2{\hat m}_1 \int \frac{d^2q_\perp dp_{2+}\phi^2 M/P_+}{(2\pi)^3}
[D+(M^2-{\delta m}^2 +2m_1^2/3 + 2m_2^2/3)\times \\  \nonumber
         & & ({2x_2M}{\hat m}_1 -D/{2M^2}) -{2/3}{\hat m}_1 D] 
+ [1 \Leftrightarrow 2]
\end{eqnarray}
where $x_2 = p_{2+}/P_+$ and has a range $0 \leq x_2 \leq +1$, to
ensure that the above expression reflects the `correct' relative 
positions of the various poles in the $p_{2-}$ plane. The further integrals 
may be evaluated using gaussian functions. The ground state $L=0$ 
function $\phi_0$ has the gaussian form
\begin{equation}\label{C.8}
\phi({\hat q}) = \exp{[-\frac{1}{2}{\hat q}^2 \beta^{-2}]} 
\end{equation}
where ${\hat q}^2$ = $q_\perp^2 +M^2 x^2$;  $x$= ${\hat m}_2-x_2$ and 
\begin{equation}\label{C.9}
D = 2M [m_2^2 + q_\perp^2 + M^2 x^2 -{\hat m}_2^2 M^2]
\end{equation}
The integration over $d^2 q_\perp$ in (C.6) yields
\begin{eqnarray}\label{C.10} 
N_V^{-2} &=& 2{\hat m}_1 \pi \beta^2 \int \frac{d x_2M}{(2\pi)^3} 
\exp{[-M^2x^2/\beta^2]} [D_\beta+(M^2-{\delta m}^2 +2m_1^2/3+2m_2^2/3)
\\  \nonumber
         & & \times ({2x_2M}{\hat m}_1 -D_\beta/{2M^2}) 
-{2/3}{\hat m}_1 D_\beta] + [1 \Leftrightarrow 2]
\end{eqnarray}
where $D_\beta$ is obtained from (C.9) by replacing $q_\perp^2$ 
with $\beta^2$. The final integration over $x_2$ is generally an
error function after changing the variable from $x_2$ to $x$=
${\hat m}_2-x_2$, which gives the limits of $x$ integration as
$$ -{\hat m}_1 \leq x \leq +{\hat m}_2 $$
In the simple case of equal mass kinematics, the final result for 
the normalization is
\begin{equation}\label{C.11}
N_V^{-2}(2\pi)^3 = 2 (\pi \beta^2)^{3/2} erf(M/{2\beta})     
[2D_{\beta^*}/3 +(M^2 +4m_q^2/3) (M/2 -D_{\beta^*}/{2M^2})] 
\end{equation}  
where 
$$ D_{\beta^*} = 2M[m_q^2 + 3\beta^2/2 - M^2/4 
-\frac{\sqrt{M^2\beta^2/\pi}\exp{(-M^2/4\beta^2)}}{2erf(M/2\beta)}] $$   
For the case of $\rho$, the normalization works out after substituting
for the input parameters from eqs (3.4-5) of text, as
\begin{equation}\label{C.12}
N_\rho  = \frac{4.340 GeV^{-3}}{(2\pi)^{3/2}}
\end{equation} 

\section*{Appendix D: Electroweak Consts $f_P$ And $g_V$}

\setcounter{equation}{0}
\renewcommand{\theequation}{D.\arabic{equation}}                  

In this Appendix we outline, mainly for illustration, short derivations of 
the pseudoscalar decay constant $f_P$ and the V-meson e.m. decay constant 
$g_V$, under the respective premises of $CIA$ and $CNPA$. 

\subsection*{D1  The P-Meson Weak Decay Constant $f_P$}

The general formula for $f_P$ via 2-quark loop is given by [59,23]          
\begin{equation}\label{D.1}
 f_P P_\mu = \sqrt{3} \int d^4 q Tr [\Psi_P \gamma_\mu \gamma_5]
\end{equation}      
where the factor $\sqrt{3}$ in front represents the effect of color [23].
Substituting from (3.13-14), and taking the traces, the integrand on RHS
simplifies to 

$$ "Tr" =\frac{4}{2i\pi} D({\hat q})\phi({\hat q}) N_P [1; P_n/M]
\frac{m_1 p_{2\mu} + m_2 p_{1\mu}}{\Delta_1 \Delta_2}$$

The next step lies in expressing the 4-vectors $p_{1,2}$ in the directons 
parallel and perpendicular to $P_\mu$ respectively, and noting that the
latter will not survive the $d^2 q_\perp$ integration. The parallel 
components in turn are 

$$ P_\mu [ {\hat m}_{1,2} \pm q.P P^{-2}] $$ 
 
The second term is just the longitudinal component $q_l$ = $q.P/M$
of $q_\mu$ which, in the $CIA$ [11] version of $MYTP$, directly contributes 
to the integral in (D.1) via the poles of the propagators $\Delta_{1,2}$ 
in this variable. In the $CNPA$ version [39] on the other hand, we have
$$ q.P = q.n P_n + q_n P.n $$ 
where the corresponding `pole' variable is proportional to $q_n$ (see
text). Then since $d^4q = d^3{\hat q}q_n$, see eq.(2.1), the `pole'
integration over $dq_l$ may be carried out exactly as in Appendix A and 
Sect.2 for $CIA$  and $CNPA$ respectively. Collecting the various factors and
simplifying, the result for $f_P$, eq.(D.1), may be expressed as a 3D
integral in either case. For definiteness, the $CIA$  result is
\begin{eqnarray}\label{D.2}
f_P &=& \frac{\sqrt{3}N_P}{(2\pi)^{3/2}} \int d^3{\phi}[2 m_{12}
(1-\frac{{\delta m}^2}{M^2})   \\  \nonumber
    & & +  \frac{{\delta m}D}{M^2}(\omega_2^{-1}- \omega_1^{-1})]
\end{eqnarray}        
where ${\delta m}=m_1-m_2$ amd $m_{12}= m_1+m_2$, and the other symbols
are as defined in Appendix A. And the 3D integration
over $d^3{\hat q}$ is a simple gaussian with the necessary substitutions
for $\phi$ and $D$-functions from text. The general
formula for unequal mass kinematics becomes finally
\begin{eqnarray}\label{D.3}
f_P &=& \sqrt{3}N_P \beta^3[2 m_{12}(1-\frac{{\delta m}^2}
{M^2}) +\frac{2{\delta m}}{M^2}<M_\omega> \\  \nonumber 
    & & \times (<\omega_2^{-1}>- <\omega_1^{-1}>)(3\beta^2 - \frac{\lambda}
{4 M^2})]
\end{eqnarray}        
A very similar result obtains for the $CNPA$ version which we
state without proof:
\begin{eqnarray}\label{D.4}
f_P &=& 2\sqrt{3}N_P \frac{\pi\beta^2}{(2\pi)^{3/2}} m_{12}
(1-\frac{{\delta m}^2}{M^2}) \\  \nonumber 
    & & \int_{-{\hat m}_1}^{+{\hat m}_2} Mdx \exp{[-M^2 x^2/{2\beta^2}]}  
\end{eqnarray}
For equal mass kinematics, the formula simplifies to
\begin{equation}\label{D.5}
f_P = 4 m_q N_P \beta^3 erf (\frac{M}{2\beta \sqrt{2}})
\end{equation}      
The physics of such quantities is discussed elsewhere [23]. We note
in passing that the value of $f_\pi$ works out as $112 MeV$, c.f. 
$133 MeV$ [23] for `half-off-shell' wave function  which agrees
with experiment [13], except that the latter does not conform to
the `angular condition' [35,9].       

\subsection*{D2  The V-Meson E.M. Decay Constant $g_V$}

In a similar way, a general formula for the e.m. decay constant of
a V-meson via 2-quark loop, is  [59, 23]
\begin{equation}\label{D.6}
\frac{M_V^2}{g_V}\epsilon_\mu  = \sqrt{3} e_Q \int d^4 q 
Tr [\Psi_V i\gamma_\mu]
\end{equation}
where $e_Q$ is the `charge' of the composite [23, 59]:
$$ e_Q^2 = \frac{1}{2} (\rho); \quad \frac{1}{18} (\omega); \quad
\frac{1}{9} (\phi); \quad \frac {4}{9}(J/psi); \quad 
\frac{1}{9}(\upsilon)$$
and the quark masses are necessarily equal. In this case, the integrand,
after taking the traces, becomes

$$ "Tr"  =\frac{4}{2i\pi} D({\hat q})\phi({\hat q}) N_V [1; P_n/M]
\epsilon_\mu \frac{(m_q^2 -p_1.p_2) + 2p_1.\epsilon p_2.\epsilon}
{\Delta_1 \Delta_2}$$

where we have anticipated that the surviving components of $p_{1,2}$
are in the direction of the polarization vector $\epsilon_\mu$ of the 
V-meson. This time let us evaluate this quantity under the $CNPA$ [39]
version of $MYTP$, following the techniques of Sect.2 of text.  
We first note the kinematical results 
$$ m_q^2-p_1.p_2 = \frac{1}{2}[\Delta_1 +\Delta_2 +M^2] ;$$
$$ 2p_1.\epsilon p_2.\epsilon \Rightarrow \frac{2}{3}p_1.p_2. $$ 
Next the $\Delta_{1,2}$ terms resulting from these reductions will
give zero contributions to the pole integration, since their poles lie
on opposite sides of the $q_n$-plane;(see Sect.2). Thus the surviving 
terms in the numerator of "Tr" are independent of $\Delta_{1,2}$. Hence 

$$ "Tr" \Rightarrow \frac{4}{2i\pi}D \phi N_V [P_n/M] \epsilon_\mu
  \frac{M^2/6 + 2 m_q^2/3}{\Delta_1 \Delta_2}$$
Recall also the $CNPA$ analogue of the $CIA$ result (A.7) of text:
$$ \frac{1}{2i\pi}\int dq_n \frac{D({\hat q}}{\Delta_1 \Delta_2}= 1 $$
which reduces the RHS of (D.6) to a trivial 3D integral, resulting in
\begin{equation}\label{D.7}
\frac{M_V^2}{g_V} = 4\sqrt{3}N_V [P_n/M_V](M^2/6 + 2 m_q^2/3)
(2\pi \beta^2)^{3/2}
\end{equation}
For a discussion of the `physics', see [23].

\end{document}